\documentclass[flushrt,preprint]{aastex}
\usepackage{graphicx}
\usepackage{url}
\usepackage{amsmath,amssymb}
\usepackage{natbib}
\begin{document}
\sloppy

\title{Long Term Trends in Atmospheric Pressure and its Variance}
\shorttitle{Atmospheric Pressure and Variance}
\shortauthors{Howells \& Katz}
\author{T. A. Howells}
\affil{Department of Physics, Washington University, St. Louis, Mo. 63130}
\author{J.~I.~Katz}
\affil{Department of Physics and McDonnell Center for the Space Sciences,\\
Washington University, St. Louis, Mo. 63130}
\affil{Tel.: 314-935-6276; Facs: 314-935-6219}
\email{katz@wuphys.wustl.edu}
\begin{abstract}
	We use the Global Historical Climatology Network--daily database to
	calculate trends in sea-level atmospheric pressures, their variance
	and the variance of their day-to-day differences in nine regions of
	the world.  Changes in pressure reflect the addition of water vapor
	to the warming atmosphere and changes in circulation patterns.
	Pressure gradients drive fronts and storm systems, and pressure
	differences, a meteorological parameter distinct from temperature
	and precipitation, are a proxy for storminess.  In eight of nine
	regions the mean sea level pressure decreased at a rate significant
	at the $2\sigma$ (95\% confidence) level if correlations between
	stations are small, but this nominal assumption is uncertain.  We
	find lower bounds on the characteristic time scale of change of the
	sea level pressure variance and its differences between consecutive
	days.  Depending on assumptions about the uncertainties of the mean
	values of trends averaged over many ($> 1000$ in some regions)
	stations, these lower bounds on the time scales of change of the
	variances range from $\sim 100$ to several thousand years.  Trends
	in the variance of day-to-day pressure differences are negative and
	nominally significant in six of nine regions.  Nominally significant
	trends in the pressure variances themselves are positive in three
	regions and negative in one.  
\end{abstract}

\section{Introduction}
Climate change is most often described by trends in temperatures or
precipitation.  These are easily measured and data are available at a
large number of sites, spread over the world.  At many sites data series
extend over a century.  There is an extensive literature, reviewed and
summarized by \cite{IPCC5AR}.

Atmospheric pressure is also readily measured, and data are available from
lengthy time series at a large number of stations.  Trends in atmospheric
pressure and its fluctuations reflect the evolving climate
\citep{GZWS03,GAA05,VW05,H06,GS09,WZSF09,BHST11,GFP13,YRHW14}.  Spatial
gradients in pressure drive winds, and their magnitude is related to the
strength of weather systems.

Compared to temperature and precipitation, sea-level adjusted pressure is
insensitive to micrometeorology, making it a robust indicator.  Changing
the ground surface around a weather station on scales of even a few meters
can change the temperature by several degrees (the urban heat island effect
is an example of this on somewhat larger scales), and even in uniform flat
terrain precipitation can be structured on scales of tens of meters or even
less.  Atmospheric pressure, averaged over gusts, is generally more uniform;
exceptions occur in intense storms where the pressure gradient structure
itself is of interest because it drives storm intensity.  Temporal structure
on a scale of days corresponds to spatial structure on a scale of 1000--2000
km, and drives cyclonic circulation.
\section{Methods}
The mean atmospheric sea level pressure $P$ is nearly constant because it is
the weight of the atmosphere, per unit area, with corrections resulting from
changes in the mean atmospheric circulation.  Several small effects change
the gravitational contribution to the mean $P$ as climate changes:
\begin{equation}
	\label{deltaP}
	{\Delta P \over P} \sim - 2 {\Delta T \over T}{h \over R}
	+ \Delta A_{\text{CO}_2} {\mu_{\text{CO}_2} - \mu_{\text{O}_2}
	\over \langle \mu_{atm} \rangle}
	+ {\mu_{\text H_2 O} \over \langle \mu_{atm} \rangle}
	{\Delta P_{\text H_2 O} \over P},
\end{equation}
where $h$ is the atmospheric scale height, $R$ the radius of the Earth, the
various $\mu$ the molecular weights of the corresponding species, $\Delta
A_{\text{CO}_2}$ the change in abundance of CO$_2$ and $\Delta P_{\text H_2
O}$ the increase in partial pressure of H$_2$O \citep{TS05}.  The first term
results from the increase in scale height as the atmosphere warms and the
decrease of gravitational acceleration with altitude, the second term the
replacement of O$_2$ by heavier CO$_2$, and the third term the increase in
atmospheric mass as the quantity of water vapor in it increases.  For a
temperature increase of $1^{\,\circ}$C, the addition of 125 ppm of CO$_2$ to
the atmosphere and an increase in water vapor content corresponding to that
temperature increase, assuming 100\% relative humidity (a crude assumption,
but sufficient to estimate the size of the effect), the first term is
about $- 7 \times 10^{-6}$, the second term is about $5 \times 10^{-5}$ and
the third term is about $7 \times 10^{-4}$, each over the roughly 150 years
since anthropogenic CO$_2$ and warming have become significant.  We might
hope that the third, and dominant, term would give an integral measurement
of the mass of water vapor in the atmosphere that is not easy to obtain in
any other manner.

In addition to a steady increase of the mean $P$ produced by the increasing
mass of the atmosphere (and very slightly offset by its thermal expansion),
changes in circulation and weather lead to regional differences in any trend
of $P$ and in its fluctuations.  These fluctuations reflect weather patterns
and their changes; storms are associated with large spatial and rapid
temporal changes in pressure.  Changes in climate, such as increased or
decreased storminess, may be reflected in changes in the variance of $P$ and
in its time dependence.

Most past studies of atmospheric pressure fluctuation \citep{RS75,CE87,GD05,
NS09,KS11} have not searched for long term trends in their statistics
that might be related to climate change.  Trends in the statistics of
pressure variations are diagnostic of climate change effects other than
warming itself.  Qualitatively, we may expect that greater pressure
variability would be correlated with more intense storms, and {\it vice
versa\/}.

We use the data in the Global Historical Climatology Network--Daily Database
\citep{MDVGH12,GHCN} as the source of pressure data.  In addition to trends
in the pressure itself, we calculate the variance of daily atmospheric
pressure measurements at station $i$ over a period denoted by $\tau$
\begin{equation}
	\mathrm{Var}[P_i]_\tau = {1 \over N_{i,\tau}} \sum\limits_{j \in \tau}
	\left(P_{i,j} - \langle P \rangle_i \right)^2,
\end{equation}
where $P_{i,j}$ is the pressure measured at station $i$ on day $j \in \tau$,
$N_i$ the number of days in $\tau$ with pressure data for station $i$ and
$\langle P \rangle_i$ is the average over all such days at station $i$.

We also define the variance of the day-to-day pressure differences
\begin{equation}
	\mathrm{Var}[\delta P_i]_\tau = {1 \over N_{i,\delta,\tau}}
	\sum\limits_{j \in \tau} \left(P_{i,j+1}-P_{i,j}\right)^2,
\end{equation}
where the sum is restricted to values of $j$ for which there are data for
both day $j$ and day $j+1$.  $N_{i,\delta,\tau} < N_{i,\tau}$, although the
difference is small because the data comprise uninterrupted runs of $n \gg 1$
days in which there are $n-1$ valid pairs.

\begin{figure}
	\centering
	\includegraphics[width=6.5in]{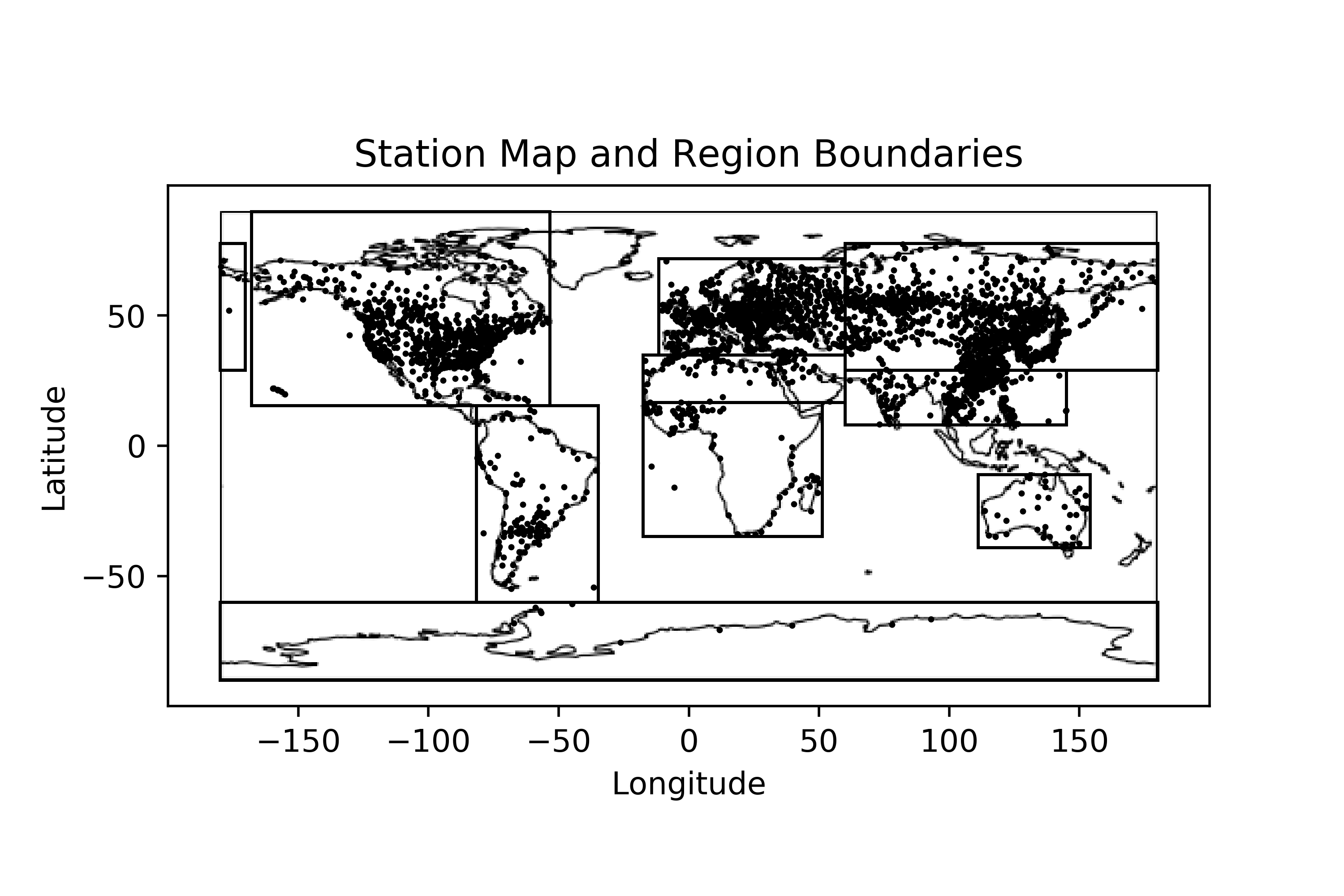}
	\caption{\label{map} Locations of sites passing tests for sufficient
	data, and the nine geographic regions over which we construct
	averages.  The North Asia region includes areas on both sides of
	longitude 180, displayed as separate boxes in this figure.}
\end{figure}

As in most historic meteorological databases, the time series in \cite{GHCN}
are incomplete.  Different stations have data from different periods of
time.  We only consider a year of data from a site to be valid if it
contains at least 150 days of data (or 150 pairs of consecutive days with
data if we are considering day-to-day pressure differences).  We divide the
period 1930--2017 into eight 11-year ``decades'', choosing 11 year intervals
so that if the Solar cycle has a significant effect on meteorological
variables each ``decade'' will sample one entire cycle; then any Solar
effect will not be aliased into a spurious long-term trend.  We consider a
``decade'' to have valid data if it contains six or more valid years and
only consider a station if it has at least four valid ``decades''.  The
stations passing this test are shown in Fig.~\ref{map}.  For these sites we
calculate the mean $P$, the mean $\mathrm{Var}[P]$ and the mean
$\mathrm{Var} [\delta P]$ for each ``decade'' and fit linear trends to these
``decadal'' means.  After fitting $dP_i/dt$, $d\mathrm{Var}[P_i]/dt$ and
$d\mathrm{Var}[\delta P_i]/dt$ at each station $i$, we average these values
over all stations in each of the nine geographic regions shown in
Fig.~\ref{map} and defined in Table~\ref{regions}.

\section{Results}
The numbers $N$ of stations passing the tests for sufficient data and the
their mean values of pressure, its variance and the variance of the
day-to-day pressure differences are shown in Table~\ref{means}.
\subsection{Mean Pressures}
\label{meanpressure}
\begin{figure}
	\centering
	\includegraphics[width=4in]{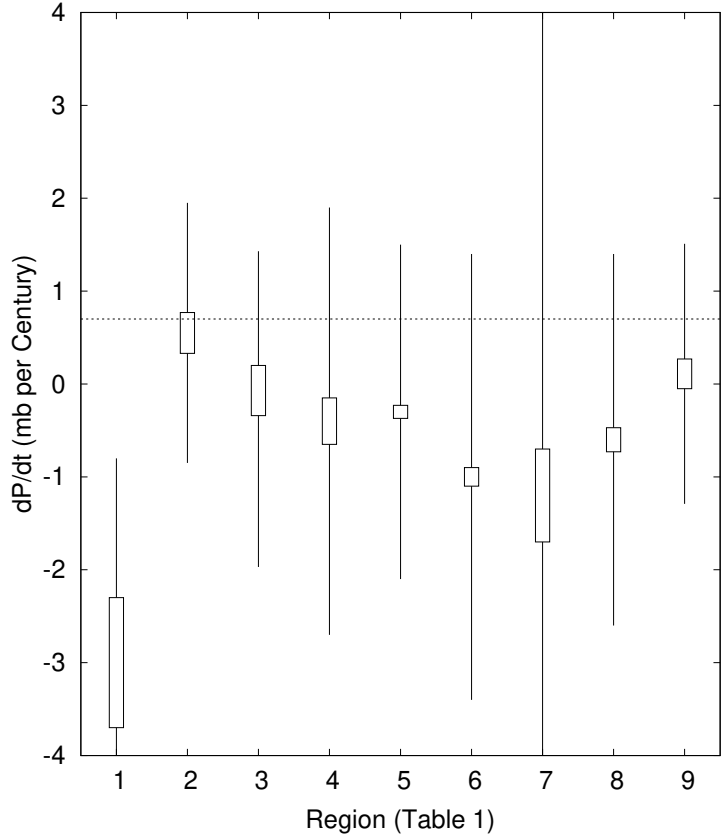}
	\caption{\label{Ptrend} Time derivatives of the mean pressure in
	each of the regions.  The boxes are $\pm 1\sigma$ uncertainties
	obtained from the standard deviations of the individual slopes,
	assuming (unrealistically) the stations are uncorrelated.  The
	error bars indicate the standard deviations of the individual
	station slopes in each region.  The horizontal line is the predicted
	slope from the increase of water vapor pressure in the atmosphere
	at a warming rate of $1^\circ$C/Century, assuming 100\% relative
	humidity and an isothermal atmosphere at $15^\circ$C.}
\end{figure}

\begin{figure}
	\centering
	\includegraphics[width=6.5in]{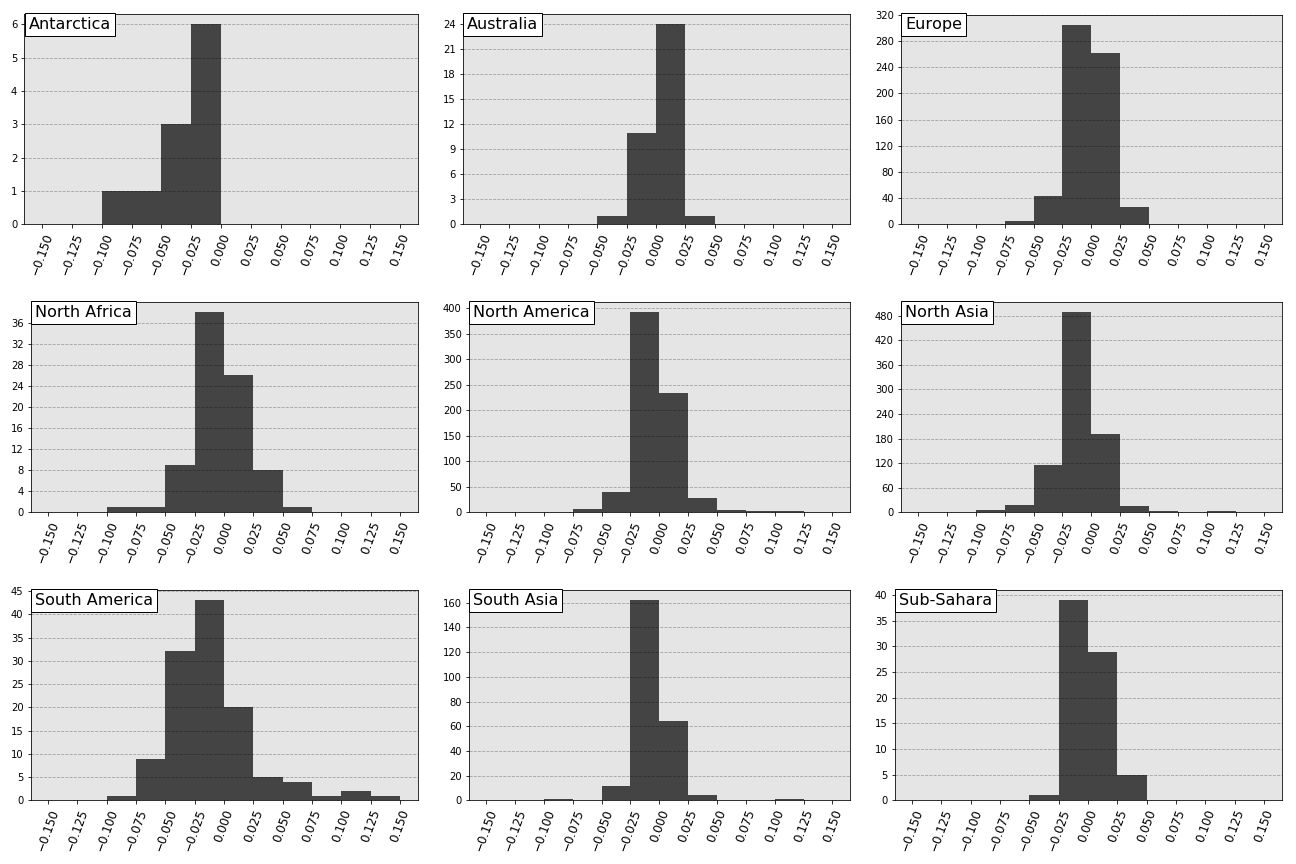}
	\caption{\label{meangrid} Distribution of fitted trends of mean
	pressure in each of the nine geographic regions in mb/year.}
\end{figure}

Fig.~\ref{Ptrend} shows the mean rates of change of pressure in each of the
nine regions defined in Table~\ref{regions}.  The boxes are the $\pm 1\sigma$
uncertainties where $\sigma$ is the standard error of the means of each
regions, and the error bars are the $\pm 1\sigma$ uncertainties where
$\sigma$ is the standard deviation of the individual station slopes; these
differ by a factor of $N^{1/2}$ where $N$ (Table~\ref{results}) is the
number of stations in each region.  The former, smaller, uncertainty would
be applicable if pressure trends at each station were independent.  In fact,
they are correlated (to an unknown degree) and the uncertainty is larger,
with the standard deviation of the individual slopes setting an upper limit.
Fig.~\ref{meangrid} shows the distributions of individual station slopes
within each region.

Comparison of the trend in mean pressure indicated by Eq.~\ref{deltaP} to
the empirical values of Table~\ref{results} shows that the predicted trend
of pressure increasing by roughly 0.5 mbar/Century is opposite in sign to
the observed (not necessarily significantly different from zero) trends,
and not insigificant in magnitude.  If the observed trends are significant
they must be attributed to changing circulation patterns and wind speeds.
Evidence for a worldwide stilling of wind speeds was reviewed by
\cite{McV12}.

The strong negative trend in Antarctic pressure may be associated with an
increasing strength of the Southern Annular Mode \citep{M03}.  \cite{McV12}
only included two Antarctic sites, at both of which wind speeds increased.
\subsection{Variances}
We calculate the mean rates of change of the variance of the
daily atmospheric pressure measurements $\mathrm{Var}[P]$ and of the
variance of the day-to-day differences of daily atmospheric pressure
measurements $\mathrm{Var}[\delta P]$ in nine regions into which we divide
the Earth's land surface.  These both measure the strength of the forces
driving weather systems.  $\mathrm{Var}[P]$ describes the overall range of
the pressure, whose gradient drives airflow, while $\mathrm{Var}[\delta P]$
provides a higher-pass filter by describing the variations on comparatively
short (one day) time scales, and therefore on shorter spatial scales as
meteorological structures advect.

\begin{figure}
	\centering
	\includegraphics[width=0.49\columnwidth]{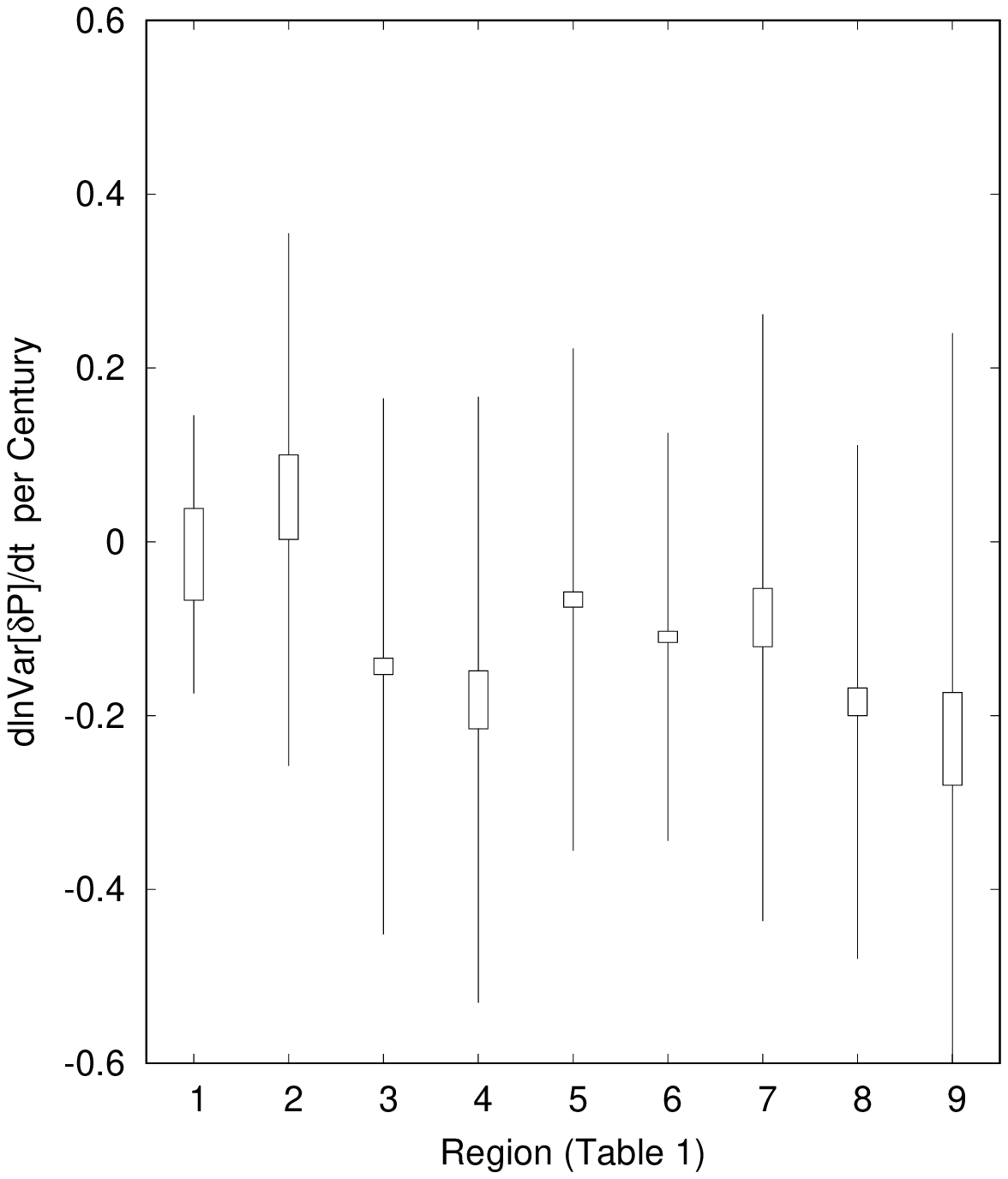}
	\includegraphics[width=0.49\columnwidth]{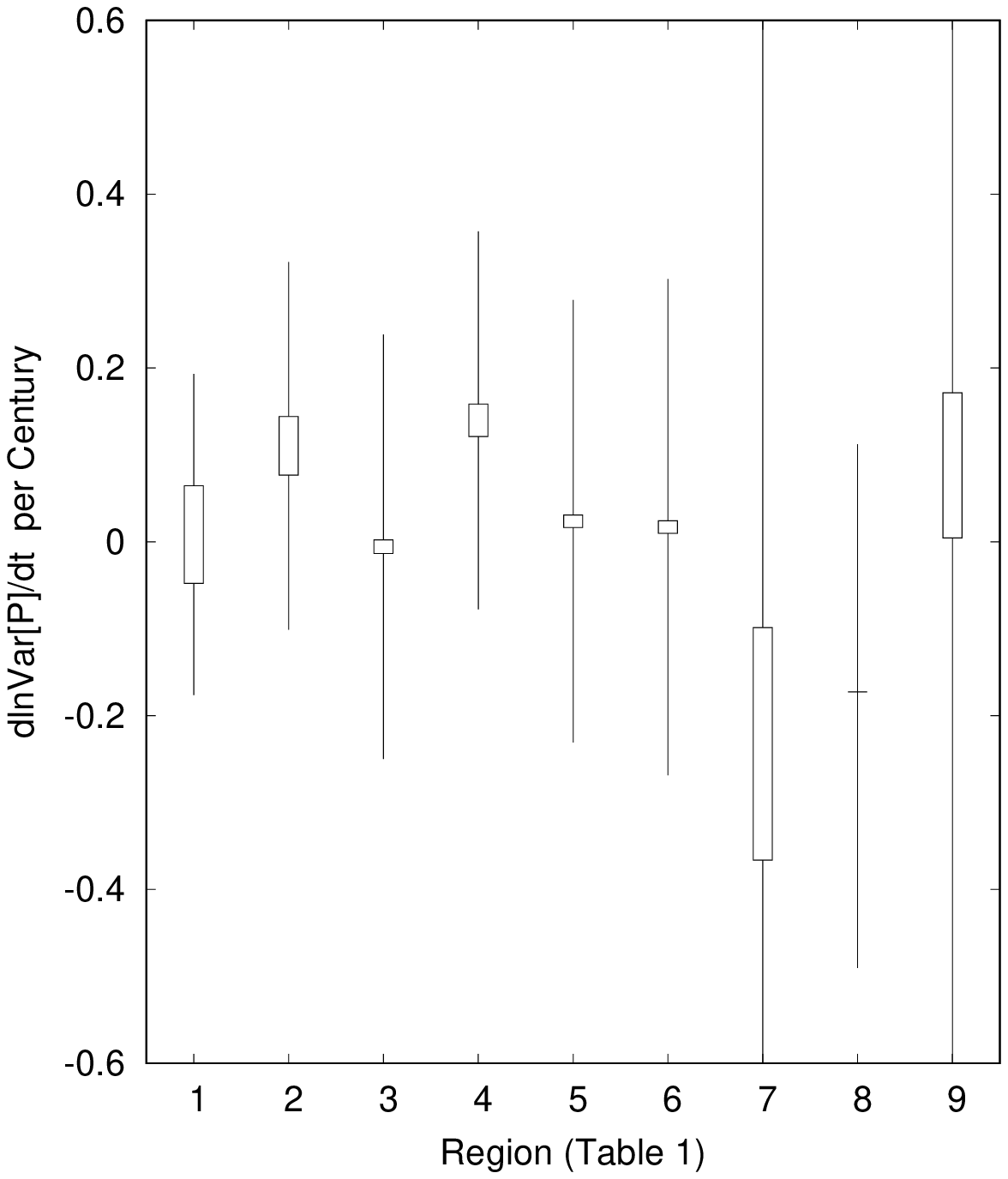}
	\caption{\label{candlesticks} The mean rates of change and $1\sigma$
	uncertainties of (left) Var[$\delta P$] and (right) Var[$P$] in each
	region.  The narrow boxes show the nominal mean errors of the means,
	assuming pressures at the stations in each region are uncorrelated,
	and the bars show the standard deviation of the rates of change at
	each site in the regions.  If the former uncertainties are adopted
	in regions 3, 4, 5, 6, 8 and 9 Var[$\delta P$] has significantly
	($> 3 \sigma$) decreasing variance, in regions 2, 4 and 5 Var[$P$]
	has significantly increasing variance and in region 8 Var[$P$] has
	significantly decreasing variance.  If the latter uncertainties are
	adopted all rates of change are consistent with zero.}
\end{figure}

\begin{figure}
	\centering
	\includegraphics[width=6.5in]{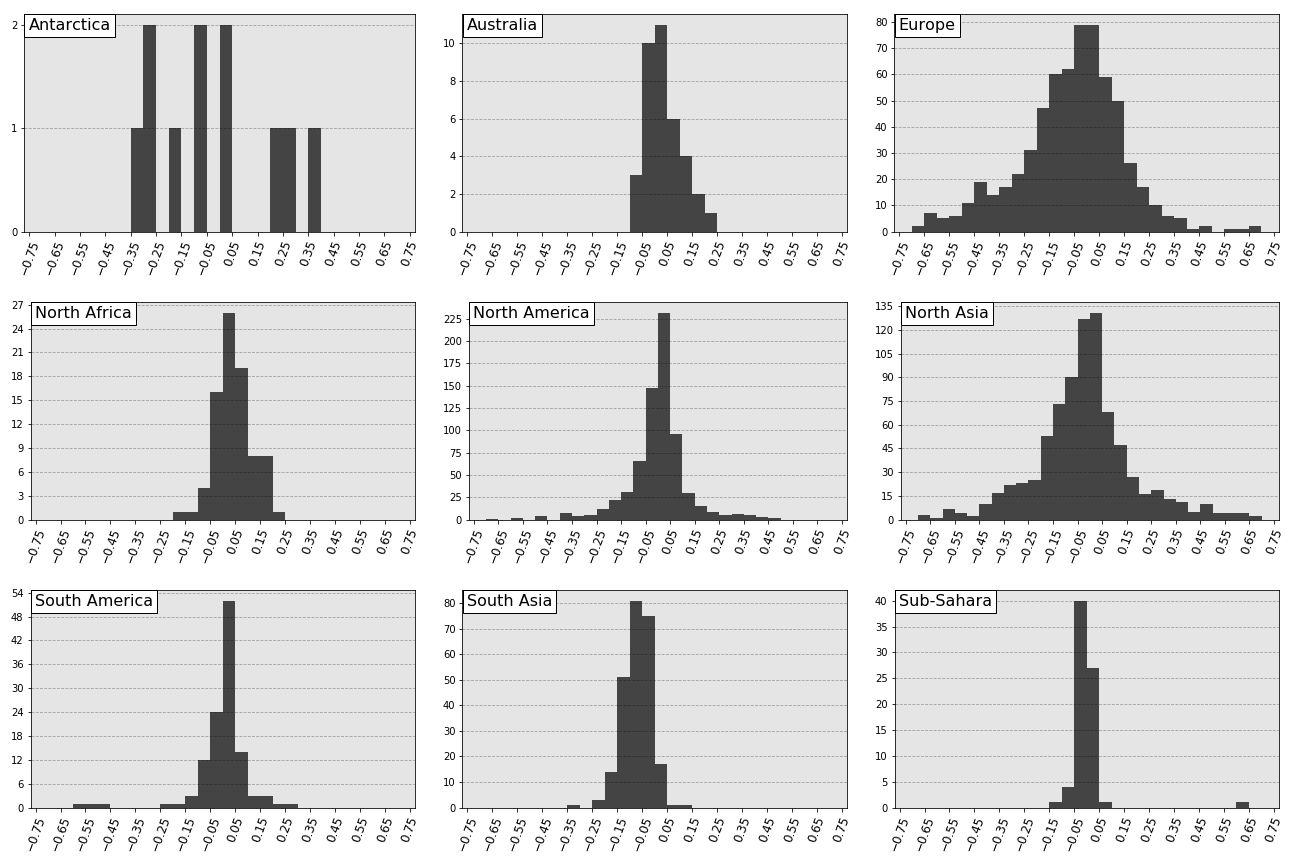}
	\caption{\label{absolute} Distribution of fitted slopes of
	$\mathrm{Var}[P]$, in \%/y, in each of the nine geographic
	regions.}
\end{figure}

\begin{figure}
	\centering
	\includegraphics[width=6.5in]{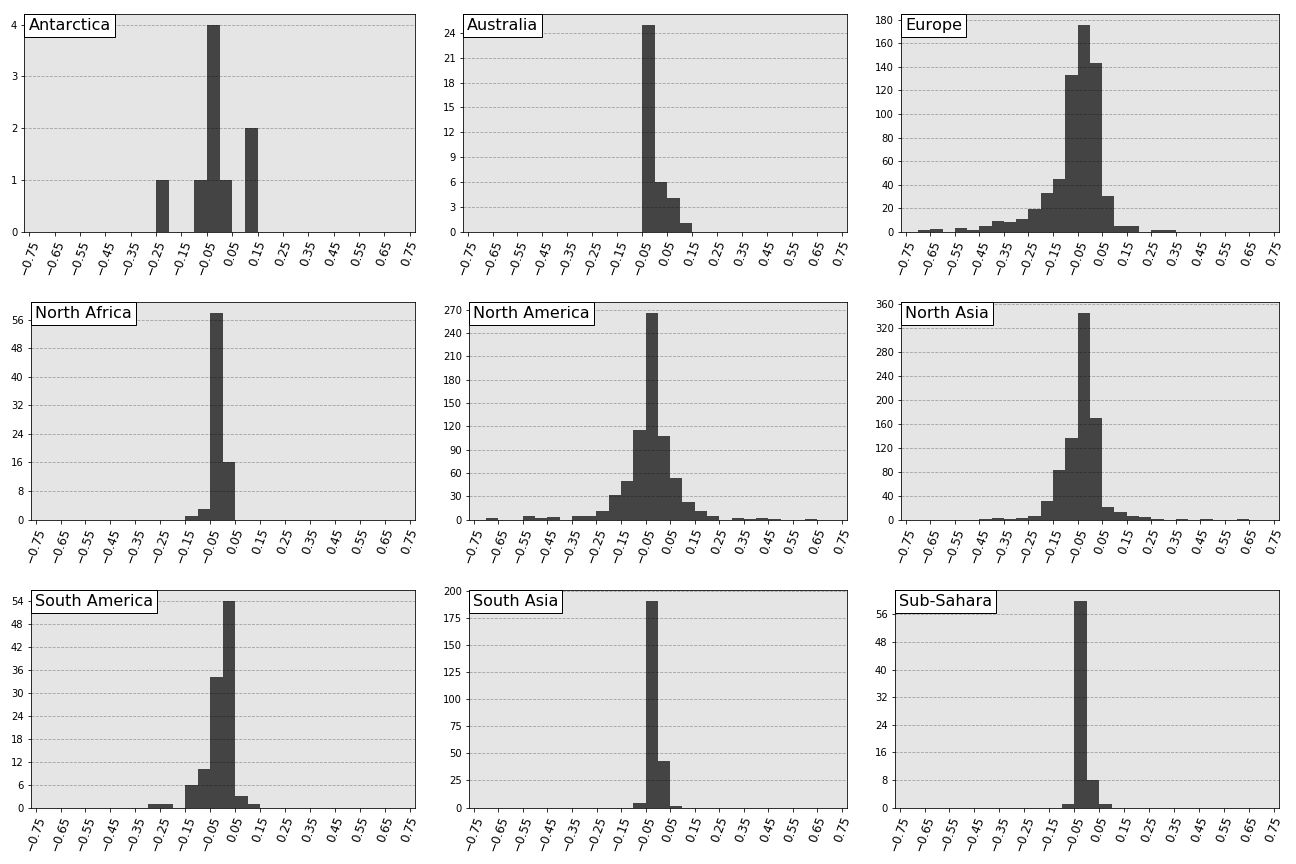}
	\caption{\label{daytoday} Distribution of fitted slopes of
	$\mathrm{Var}[\delta P]$, where $\delta P$ is the pressure
	difference between two consecutive days, in \%/y, in each of
	the nine geographic regions.}
\end{figure}

Table~\ref{regions} defines the regions and also gives lower bounds on the
characteristic times of pressure variance change (the variances divided by
their time derivatives) for each region and for both variances.  The
numerical results are shown in Table~\ref{results} and are plotted in
Fig.~\ref{candlesticks}.  The distributions of fitted slopes in each region
are shown in Figs.~\ref{absolute} and \ref{daytoday}.

\section{Discussion}
The decrease of mean pressure in most regions is opposite to the effect of
adding additional H$_2$O vapor to the atmosphere.  The fact that it is
observed in eight of the nine regions, and statistically significant at the
$2\sigma$ level in seven of them if the smaller of the uncertainty estimates
is valid, suggests that it may be real.  It must be attributed to changes in
the atmospheric flow.

If the smaller (standard error of the means) uncertainty estimates are 
adopted then in six regions a significant decrease in day-to-day pressure
variance, at a rate $\sim 10^{-3}$/y, is found (Fig.~\ref{candlesticks}).
If the larger uncertainty estimates are adopted then no significant trend is
found, but an upper limit to any rate of change of these variances $\sim
0.5\text{--}1 \times 10^{-2}$/y can be set in most regions
(Table~\ref{regions}).  The values in this Table are all lower bounds on
$T_{char}$ (upper bounds on rates of change), so outlying small values
indicate only that data in those regions are scattered, not that the
variances are changing rapidly.

These conclusions are similar to the earlier result \citep{LCK} that
storminess in the 48 contiguous United States, as measured by the normalized
variance of hourly rainfall, has shown little or no trend in a period that
substantially overlaps with that of this study.  Increasing temperature has
not been associated with rapid change in the other meteorological parameters
studied.  
\clearpage
\bibliography{atmpressijoc}

\begin{thebibliography}{}
\expandafter\ifx\csname natexlab\endcsname\relax\def\natexlab#1{#1}\fi

\bibitem[{Canavero \& Einaudi(1987)}]{CE87}
Canavero, F.~G., \& Einaudi, F. 1987, Time and Space Variability of Spectral
  Estimates of Atmospheric Pressure, J. Atm. Sci., 44, 1589

\bibitem[{Canel \& Katz(2018)}]{LCK}
Canel, L.~M., \& Katz, J.~I. 2018, Trends in U. S. Hourly Precipitation
  Variance 1949--2009, J. Hydromet., 19, 599

\bibitem[{Finkel \& Katz(2018)}]{FK18}
Finkel, J.~M., \& Katz, J.~I. 2018, Changing World Extreme Temperature
  Statistics, Int. J. Clim., 38, 2613

\bibitem[{GHCN(2018)}]{GHCN}
GHCN. 2018, \url{ftp://ftp.ncdc.noaa.gov/pub/data/ghcn/daily}, accessed
  7/9/2018

\bibitem[{Gillett {et~al.}(2005)Gillett, Allan, \& Ansell}]{GAA05}
Gillett, N.~P., Allan, R.~J., \& Ansell, T.~J. 2005, Detection of external
  influence on sea level pressure with a multi-modal ensemble, Geophys. Res.
  Lett., 32, L19714

\bibitem[{Gillett {et~al.}(2013)Gillett, Fyfe, \& Parker}]{GFP13}
Gillett, N.~P., Fyfe, J.~C., \& Parker, D.~E. 2013, Attribution of observed sea
  level pressure trends to greenhouse gas, aerosol and ozone changes, Geophys.
  Res. Lett., 40, 2302

\bibitem[{Gillett \& Stott(2009)}]{GS09}
Gillett, N.~P., \& Stott, P.~A. 2009, Attribution of anthropogenic influence on
  seasonal sea level pressure, Geophys. Res. Lett., 36, L23709

\bibitem[{Gillett {et~al.}(2003)Gillett, Zwiers, Weaver, \& Stott}]{GZWS03}
Gillett, N.~P., Zwiers, F.~W., Weaver, A.~J., \& Stott, P.~A. 2003, Detection
  of human influence on sea-level pressure, Nature, 422, 292

\bibitem[{Gong \& Drange(2005)}]{GD05}
Gong, D., \& Drange, H. 2005, A Preliminary Study on the Relationship Between
  Arctic Oscillation and Daily SLP Variance in teh Northern Hemisphere During
  Wintertime, Adv. Atm. Sci., 22, 313

\bibitem[{Hegerl {et~al.}(2006)Hegerl, Karl, Allen, Bindoff, Gilett, Karoly,
  Zhang, \& Zwiers}]{H06}
Hegerl, G.~C., Karl, T.~R., Allen, M., {et~al.} 2006, Climate change detection
  and attribution: Beyond mean temperature signals, J. Clim., 19, 5058

\bibitem[{{Intergovernmental Panel on Climate Change}(2013--2014)}]{IPCC5AR}
{Intergovernmental Panel on Climate Change}. 2013--2014, IPCC Fifth Assessment
  Report (IPCC), \url{www.ipcc.ch/report/ar5/index.shtml}

\bibitem[{Krueger \& {von Storch}(2011)}]{KS11}
Krueger, O., \& {von Storch}, H. 2011, Evaluation of an Air Pressure Based
  Proxy for Storm Activity, J. Clim., 24, 2612

\bibitem[{Marshall(2003)}]{M03}
Marshall, G.~J. 2003, Trends in the Southern Annular Mode from Observations and
  Reanalyses, J. Clim., 16, 4134

\bibitem[{McVicar {et~al.}(2012)McVicar, Roderick, Donohue, Li, Niel,
  ad~J.~Grieser, Jhajharia, Himri, Mahowald, Mescherskaya, Kruger, Rehman, \&
  Dinpashoh}]{McV12}
McVicar, T.~R., Roderick, M.~L., Donohue, R.~J., {et~al.} 2012, Global review
  and synthesis of trends in observed terrestrial near-surface wind speeds:
  Implications for evaporation, J. Hydrology, 416--417, 182

\bibitem[{Menne {et~al.}(2012)Menne, Durre, Vose, Gleason, \&
  Houston}]{MDVGH12}
Menne, J.~M., Durre, I., Vose, R.~S., Gleason, B.~E., \& Houston, T.~G. 2012,
  An overview of the global historical climatology network---daily database, J.
  Atmos. Ocean. Tech., 29, 897, doi:10.1175/JTECH-D-11-00103.1

\bibitem[{Nawri \& Stewart(2009)}]{NS09}
Nawri, N., \& Stewart, R.~E. 2009, Short-term temporal variability of
  atmospheric surface pressure an dwind speed in the Canadian Arctic, Theor.
  Appl. Climatol., 98, 151

\bibitem[{Rosenan \& Striem(1975)}]{RS75}
Rosenan, N., \& Striem, H.~L. 1975, The Mean Daily Variation of Barometric
  Pressure, Its Characteristics and its Its Contituents, in Israel and
  Neighbouring Countries, Arch. Met. Geoph. Biokl., Ser. A, 24, 329

\bibitem[{Trenberth \& Smith(2005)}]{TS05}
Trenberth, K., \& Smith, L. 2005, The mass of the atmosphere: A constraint on
  global analyses, J. Climate, 18, 864

\bibitem[{{van den Besselaar} {et~al.}(2011){van den Besselaar}, Haylock, {van
  der Schrier}, \& Tank}]{BHST11}
{van den Besselaar}, E. J.~M., Haylock, M.~R., {van der Schrier}, G., \& Tank,
  A. M. G.~K. 2011, A European daily high-resolution observational gridded data
  set of sea level pressure, J. Geophys. Res.--Atm., 116, D11110

\bibitem[{{Van Wijngaarden}(2005)}]{VW05}
{Van Wijngaarden}, W.~A. 2005, Examination of trends in hourly surface pressure
  in Canada during 1953--2003, Int. J. Clim., 25, 2041

\bibitem[{Wang {et~al.}(2009)Wang, Zwiers, Swail, \& Feng}]{WZSF09}
Wang, X. L.~L., Zwiers, F.~W., Swail, V.~R., \& Feng, Y. 2009, Trends and
  variability of storminess in the Northeast Atlantic region, 1874--2007, Clim.
  Dyn., 33, 1179

\bibitem[{Yu {et~al.}(2014)Yu, Ren, Hu, \& Wu}]{YRHW14}
Yu, Y.~Y., Ren, R.~C., Hu, J.~G., \& Wu, G.~X. 2014, A Mass Budget Analysis on
  the Interannual Variability of the Polar Surface Pressure in the Winter
  Season, J. Atm. Sci., 71, 3539

\end{thebibliography}
\begin{table}
	\centering
	\begin{tabular}{|c|l|r|cc|rr|rr|}
		\hline
		\# & Region & $N$ & Latitudes & Longitudes & \multicolumn{2}{c|}
		{$T_{char,\delta P}$ (y)} & \multicolumn{2}{c|}
		{$T_{char,P}$ (y)} \\
		& & & & & s.d. & s.e. & s.d. & s.e. \\
		\hline
		1 & Antarctica & 9 & 60--90 S & All & 299 & 834 & 265 & 827 \\
		2 & Australia & 36 & 10.93--39.1 S & 111--154 E & 151 & 660 & 187 & 562 \\
		3 & Europe & 984 & 35--72 N & 11.55 W--60 E & 131 & 617 & 203 & 4770 \\
		4 & N. Africa{/ME} & 106 & 16.7--35 N & 17.7 W--60 E & 113 & 396 & 174 & 565 \\
		5 & N. America & 1095 & 15.5--90 N & 53.5--168 W & 155 & 1190 & 188 & 2620 \\
		6 & N. Asia & 1438 & 29--77.7 N & 60 E--170.4 W & 173 & 818 & 170 & 3200 \\
		7 & S. America & 120 & 15.5 N--60 S & 34.9--81.7 W & 128 & 650 & 30 & 200 \\
		8 & S. Asia & 326 & 8.15--29 N & 60--145 E & 127 & 462 & 126 & 451 \\
		9 & Sub-Sahara & 78 & 34.8 S--16.7 N & 17.7 W--51.2 E & 86 & 298 & 60 & 391 \\
		\hline
	\end{tabular}
	\caption{\label{regions} The nine regions, as defined in
	\cite{FK18}.  $N$ is the number of stations in each region with
	sufficient data to fit slopes to the time dependence of the
	variances of the day-to-day pressure differences.  The criteria are
	at least 150 pairs of consecutive days in a year, at least six such
	years in an 11 year ``decade'', and at least four such ``decades''
	out of the eight from 1930--2017.  A few more stations meet the
	criteria for the variances of the pressure $P$ that do not require
	consecutive pairs of days.  The characteristic times $T_{char}$ (in
	years), separately for $\delta P$ and $P$, are defined as the ratios
	of the mean variances to the largest absolute values of their time
	derivatives within $2\sigma$ of their best fits.  Lower bounds on
	$T_{car}$ are shown for two definitions of $\sigma$: The first
	bounds use the (large) standard deviations (s.d.) of the station
	variances, and the second bounds use the standard errors (s.e.) of
	their means, smaller by $N^{1/2}$, and hence imply much larger lower
	bounds on $T_{char}$.  These are only bounds because all the mean
	slopes are consistent with zero (unless the actual uncertainties are
	nearly as small as the standard errors of their means, which would
	not be so for strongly correlated data).}
\end{table}

\begin{table}
\centering
\begin{tabular}{|l|c|c|c|}
\hline
	Region & $\langle dP/dt \rangle$ (mb/C) & 
	$\langle d{\mathrm Var}[\delta P]/dt \rangle$ (mb$^2$/C) & 
	$\langle d{\mathrm Var}[P]/dt \rangle$ (mb$^2$/C) \\
\hline
	Antarctica & $-3.0 \pm 2.2 \pm 0.7$ & $-0.9 \pm 10. \pm 3.3$ & $1.0 \pm 22. \pm 6.7$ \\
	Australia & $0.55 \pm 1.4 \pm 0.22$ & $0.54 \pm 3.4 \pm 0.57$ & $3.6 \pm 6.9 \pm 1.1$ \\
	Europe & $-0.27 \pm 1.7 \pm 0.07$ & $-4.6 \pm 9.9 \pm 0.3$ & $-0.5 \pm 22.  \pm 0.7$ \\
	N. Africa{/ME} & $-0.40 \pm 2.4 \pm 0.25$ & $-1.2 \pm 2.3 \pm 0.22$ & $4.5 \pm 7.0 \pm 0.6$ \\
	N. America & $-0.30 \pm 1.8 \pm 0.07$ & $-2.3 \pm 10. \pm 0.3$ & $1.3 \pm 14. \pm 0.4$ \\
	N. Asia & $-1.0 \pm 2.4 \pm 0.1$ & $-3.4 \pm 7.3 \pm 0.2$ & $1.9 \pm 32. \pm 0.8$ \\
	S. America & $-1.2 \pm 5.5 \pm 0.5$ & $-1.3 \pm 5.2 \pm 0.5$ & $-6.6 \pm 44. \pm 3.8$ \\
	S. Asia & $-0.60 \pm 2.0 \pm 0.13$ & $-0.81 \pm 1.3 \pm 0.07$ & $-6.9 \pm 11. \pm 0.6$ \\
	Sub-Sahara & $0.11 \pm 1.4 \pm 0.16$ & $-0.68 \pm 1.4 \pm 0.16$ & $0.80 \pm 7.2 \pm 0.76$ \\
\hline
\end{tabular}
	\caption{\label{results} The mean rates of change of the pressure in
	mb/Century, of the variance of $\delta P$ in mb$^2$/Century, and of
	the variance of $P$ in mb$^2$/Century.  Two $1\sigma$ uncertainties
	of the slopes are presented: the standard deviations of the slopes
	of the variances at the individual stations in the region, which
	overestimate the uncertainties of their means, and the standard
	errors of the mean slopes, which underestimate the uncertainties of
	their means because pressure among the stations is correlated.
	These uncertainties are not additive.}
\end{table}

\begin{table}
	\centering
	\begin{tabular}{|l|c|c|c|}
		\hline
		Region & $\langle P \rangle$ (mb) & $\langle
		\mathrm{Var}[\delta P]\rangle$ (mb$^2$) & $\langle
		\mathrm{Var}[P]\rangle$ (mb$^2$) \\
		\hline
		Antarctica & 987.9 & 62.5 & 119.1 \\
		Australia & 1014.5 & 11.1 & 32.6 \\
		Europe & 1015.6 & 32.1 & 90.1 \\
		N. Africa/ME & 1014.0 & 6.6 & 32.2 \\
		N. America & 1015.7 & 34.6 & 55. \\
		N. Asia & 1016.4 & 31.1 & 112.1 \\
		S. America & 1013.0 & 14.9 & 28.4 \\
		S. Asia & 1011.1 & 4.4 & 36.5 \\
		Sub-Sahara & 1012.2 & 3.0 & 9.1 \\
		\hline
	\end{tabular}
	\caption{\label{means} Mean values of pressure $P$, its variance and
	the variance of its differences $\delta P$ on consecutive days.}
\end{table}

\begin{acknowledgements}
	We thank J. Finkel for assistance with Fig.~\ref{map} and P. Huybers
	and W. H. Press for useful discussions.
\end{acknowledgements}
\end{document}